\documentclass[twocolumn,showpacs,superscriptaddress]{revtex4-1}

\input{epsf}
\usepackage{amssymb}

\begin{document}


\title{Modeling the nongravitational acceleration during Cassini's gravitation experiments}

\date{\today}


\author{O. Bertolami}
\email{orfeu.bertolami@fc.up.pt}
\affiliation{Departamento de F\'isica e Astronomia and Centro de F\'isica do Porto, Faculdade de Ci\^encias, Universidade do Porto, Rua do Campo Alegre 687, 4169-007 Porto, Portugal}
\author{F. Francisco}
\email{frederico.francisco@tecnico.ulisboa.pt}
\affiliation{Instituto de Plasmas e Fus\~ao Nuclear, Instituto Superior T\'ecnico, Universidade de Lisboa, Avenida Rovisco Pais 1, 1049-001 Lisboa, Portugal}
\author{P.J.S. Gil}
\email{p.gil@dem.ist.utl.pt}
\affiliation{LAETA, IDMEC, Instituto Superior T\'ecnico, Universidade de Lisboa, Avenida Rovisco Pais 1, 1049-001 Lisboa, Portugal}
\author{J. P\'aramos}
\email{jorge.paramos@fc.up.pt}
\affiliation{Departamento de F\'isica e Astronomia and Centro de F\'isica do Porto, Faculdade de Ci\^encias, Universidade do Porto, Rua do Campo Alegre 687, 4169-007 Porto, Portugal}


\begin{abstract}
	In this paper we present a computation of the thermally generated acceleration of the Cassini probe during its solar conjunction experiment, obtained from a model of the spacecraft. We build a thermal model of the vehicle and perform a Monte Carlo simulation to find a thermal acceleration with a main component of $(3.01 \pm 0.33) \times 10^{-9}~{\rm m/s^2}$. This result is in close agreement with the estimates of this effect performed through Doppler data analysis.
\end{abstract}


\maketitle


\section{Introduction}

The Cassini mission was launched on October 15th 1997. Its goal was to reach Saturn and also included a set of planned experiments designed to test General Relativity. One of these experiments was carried out from June 6th to July 7th 2002, while the probe was in a solar conjunction. The results from the data harvested during this one month period allowed for constraining the $\gamma$ parameter of the PPN formalism, which quantifies the amount of curvature generated per unit mass, to within $(2.1 \pm 2.3) \times 10^{-5}$ of unity, the most accurate bound obtained so far \cite{Bertotti:2003}.

During the solar conjunction experiment, the non-gravitational acceleration had to be filtered out as well as possible and, in particular, the significant contributions from solar radiation pressure and from anisotropic thermal emission of the probe itself. Due to the unavailability of any straightforward procedure to obtain the said thermal emission from a model of the spacecraft, data from Doppler measurements was used to estimate the component of the acceleration that is constant relative to the spacecraft orientation. The obtained values for the thermally generated acceleration reveal that the largest component is aligned with the Earth-spacecraft axis and amounts to $3 \times 10^{-9}~{\rm m/s^2}$ towards the Earth. The other two components are smaller and measured orthogonally to the orbital plane and on the orbital plane, and are found to be about $4 \times 10^{-10}~{\rm m/s^2}$ and $1 \times 10^{-10}~{\rm m/s^2}$, respectively. These components, however, have large error estimates associated with their determination \cite{Bertotti:2003}.

The aim of this paper is to consider the problem of obtaining the value of the thermally generated accelerations and directly respond to the stated difficulty in extracting them from a model of the spacecraft itself. It is shown that reliable results can be obtained by using the physical and computational framework previously developed to study the the acceleration generated by thermal emissions in the Pioneer 10 and 11 spacecraft, in the context of the problem that became known as \textit{the Pioneer anomaly} \cite{Bertolami:2008,Bertolami:2009,Francisco:2012}.


\section{Pointlike source method}


\subsection{Motivation}

The pointlike source method is an approach that maintains a high computational speed and a broad degree of flexibility, allowing for an easy analysis of different contributions and scenarios.

The method was designed to keep all the physical features of the problem at glance and all steps easy to scrutinize. Although it can be argued that this simplicity and transparency was achieved at the expense of accuracy, a battery of test cases can be performed to test the robustness of the results \cite{Bertolami:2008,Bertolami:2009}. These test cases validate the approach, as they show that, for reasonable assumptions, the possible lack of accuracy caused by our modeling approach is much smaller than the accuracy in the characterization of the acceleration itself.

This method was also designed to consider parameters involving a large degree of uncertainty: this is related to the geometrical and material properties of the various spacecraft elements, which in most cases do not have well-known baseline (before launch) values, and have endured extended periods of degradation in space. By assigning a statistical distribution to each parameter, based on the available information, and generating a large number of random values, we have used a Monte Carlo simulation to obtain a probability distribution for the final result \cite{Francisco:2012}.

The fact that this method was already used to deal with spacecraft thermal emissions in the context of the Pioneer anomaly, producing results that are generally in agreement with the ones obtained through subsequent, more detailed finite-element models \cite{Rievers:2011,Turyshev:2012}, is a further indication of its reliability and robustness.


\subsection{Radiative Momentum Transfer}
\label{Sec:RadMomTransf}

Before considering the particular problem at hand, it is useful to briefly review the physical formulation behind the pointlike source method.

The key feature of this method is a distribution of a small number of carefully placed pointlike radiation sources that models the thermal radiation emissions of the spacecraft. One typically uses Lambertian radiation sources to model surface emissions, however, other types of sources may be used to model particular objects.

All the subsequent formulation of emission and reflection is made in terms of the Poynting vector-field. For instance, the time-averaged Poynting vector-field for a Lambertian source located at $\mathbf{x}_0$ is given by
\begin{equation}
	\label{lambertian2}
	\mathbf{S}_{\text{Lamb}}(\mathbf{x})={W \over \pi ||\mathbf{x}-\mathbf{x}_0||^2} \left( \mathbf{n} \cdot {\mathbf{x}-\mathbf{x}_0 \over ||\mathbf{x}-\mathbf{x}_0||} \right) {\mathbf{x}-\mathbf{x}_0 \over ||\mathbf{x}-\mathbf{x}_0||},
\end{equation}
where $W$ is the emissive power and $\mathbf{n}$ is the surface normal.

In this work, we introduce a small extension to the pointlike source method to include other radiation source geometries, as long as they have a straightforward mathematical description. An especially useful example is the cylindrical source, where the emitter is a line segment instead of a point and the Poynting vector field has cylindrical symmetry. For instance, the radiation field of a cylindrical source parallel to the $x$-axis with coordinates $(y_0,z_0)$ in the $yz$-plane is given by
\begin{equation}
	\label{cylindrical}
	\mathbf{S}_{\text{cyl}}(\mathbf{x})={W (0,y-y_0,z-z_0) \over 2 \pi l ((y-y_0)^2 + (z-z_0)^2)},
\end{equation}
where $l$ is the length of the source and $\mathbf{x}=(x,y,z)$.

The amount of power illuminating a given surface $W_{\rm ilum}$ can be obtained by computing the Poynting vector flux through the illuminated surface $S$, given by the integral
\begin{equation}
	E_{\rm ilum} = \int_S \mathbf{S} \cdot \mathbf{n}_{\rm ilum}~ dA,
\end{equation}
where $\mathbf{n}_{\rm ilum}$ is the normal vector of the illuminated surface.

The absorbed radiation transfers its momentum to the surface yielding a \emph{radiation pressure} $p_{\rm rad}$ given, for an opaque unit surface, by the power flux divided by the speed of light. There is also a radiation pressure on the emitting surface but with its sign reversed. If there is transmission (\textit{i.e.}, the surface is not opaque) the pressure is multiplied by the absorption coefficient. As to reflection, we shall see in the next two sections that it is treated as a re-emission of a part of the absorbed radiation.

Integrating the radiation pressure on a surface allows us to obtain the force and, dividing by the mass of the spacecraft, the acceleration
\begin{equation}
	\label{force_integration}
	\mathbf{a}_{\rm th} = {1 \over m_{\rm Cassini}} \int_S {\mathbf{S} \cdot \mathbf{n}_{\rm ilum} \over c} {\mathbf{S} \over ||\mathbf{S}||} dA.
\end{equation}

The procedure to compute this integration is not always straightforward: to determine the force exerted by the radiation on the emitting surface, the integral should be taken over a closed surface encompassing the latter; equivalently, the force exerted by the radiation on an illuminated surface requires an integration surface that encompasses it. Furthermore, considering a set of emitting and illuminated surfaces implies the proper accounting of the effect of the shadows cast by the various surfaces, which are then subtracted from the estimated force on the emitting surface. One may then read the thermally induced acceleration directly,


\subsection{Reflection Modeling -- Phong Shading}
\label{Sec:Phong}

The inclusion of reflections in the model is achieved through a method known as \emph{Phong Shading}, a set of techniques and algorithms commonly used to render the illumination of surfaces in 3D computer graphics \cite{Phong:1975}.

This method is composed of a reflection model including diffusive and specular reflection, known as \emph{Phong reflection model}, and an interpolation method for curved surfaces modeled as polygons, known as \emph{Phong interpolation}.

The Phong reflection model is based on an empirical expression that gives the illumination value of a given point in a surface, $I_{\rm p}$, as
\begin{equation}
	\label{phong_refl_mod}
	I_{\rm p} = k_{\rm a} i_{\rm a} + \sum_{m \in \text{lights}} \left[k_{\rm d} (\mathbf{l}_m \cdot \mathbf{n})i_{\rm d} + k_{\rm s} (\mathbf{r}_m \cdot \mathbf{v})^{\alpha} i_{\rm s} \right],
\end{equation}
where $k_{\rm a}$, $k_{\rm d}$ and $k_{\rm s}$ are the ambient, diffusive and specular reflection constants, $i_{\rm a}$, $i_{\rm d}$ and $i_{\rm s}$ are the respective light source intensities, $\mathbf{l}_m$ is the direction of the light source $m$, $\mathbf{n}$ is the surface normal, $\mathbf{r}_m$ is the direction of the reflected ray, $\mathbf{v}$ is the direction of the observer and $\alpha$ is a ``shininess'' constant (the higher it is, the more mirror-like is the surface).

In using this formulation to resolve a physics problem, there are a few constrains that should be taken into account. The ambient light parameter $k_{\rm a}$ and $i_{\rm a}$, while useful in computer graphics, are not relevant for this problem since they give the reflection behavior relative to a background radiation source. Also, the intensities $i_{\rm d}$ and $i_{\rm s}$ should be the same, since the diffusive and specular reflection are relative to the same radiation sources.

This method provides a simple and straightforward way to model the various components of reflection, as well as a more accurate accounting of the thermal radiation exchanges between the surfaces on the spacecraft. In principle, there is no difference between the treatment of infrared radiation, in which we are interested, and visible light, for which the method was originally devised, allowing for a natural wavelength dependence of the above material constants.

Given the presentation of the thermal radiation put forward in subsection \ref{Sec:RadMomTransf}, the Phong shading methodology was adapted from a formulation based on \emph{intensities} (energy per unit surface of the projected emitting surface) to one based on the energy-flux per unit surface (the Poynting vector).


\subsection{Computation of Reflection}

Using the formulation outlined in section \ref{Sec:Phong}, the diffusive and specular components of reflection can be separately computed in terms of the Poynting vector-field. The reflected radiation Poynting vector-field for the diffusive component of the reflection is given by
\begin{equation}
	\label{diffusive_reflection}
	\mathbf{S}_{\rm rd}(\mathbf{x},\mathbf{x}')={k_{\rm d} |\mathbf{S}(\mathbf{x}')\cdot \mathbf{n}| \over \pi ||\mathbf{x}-\mathbf{x}'||^2} \mathbf{n} \cdot (\mathbf{x}-\mathbf{x}') {\mathbf{x}-\mathbf{x}' \over ||\mathbf{x}-\mathbf{x}'||},
\end{equation}
while the specular component reads

\begin{equation}
	\label{specular_reflection}
	\mathbf{S}_{\rm rs}(\mathbf{x},\mathbf{x}')={k_{\rm s} |\mathbf{S}(\mathbf{x}')\cdot \mathbf{n}| \over {2 \pi \over 1+ \alpha} ||\mathbf{x}-\mathbf{x}'||^2} [\mathbf{r} \cdot (\mathbf{x}-\mathbf{x}')]^{\alpha} {\mathbf{x}-\mathbf{x}' \over ||\mathbf{x}-\mathbf{x}'||}.
\end{equation}
where $\mathbf{x}'$ is a point on the reflecting surface. In both cases, the reflected radiation field depends on the incident radiation field $\mathbf{S}(\mathbf{x}')$ and on the reflection coefficients $k_{\rm d}$ and $k_{\rm s}$, respectively. Using Eqs.~(\ref{diffusive_reflection}) and (\ref{specular_reflection}), one can compute the reflected radiation field by adding up these diffusive and specular components. From the emitted and reflected radiation vector fields, the irradiation of each surface is computed and, from that, a calculation of the force can be performed through Eq.~(\ref{force_integration}).

In the modeling of the actual vehicle, once the radiation source distribution is put in place, the first step is to compute the emitted radiation field and the respective force exerted on the emitting surfaces. This is followed by the determination of which surfaces are illuminated and the computation of the force exerted on those surfaces by the radiation. At this stage, we get a figure for the thermal force without reflections. The reflection radiation field is then computed for each surface and subject to the same steps as the initially emitted radiation field, leading to a determination of thermal force with one reflection.

This method can, in principle, be iteratively extended to as many reflection steps as desired, considering the numerical integration algorithms and available computational power. If deemed necessary, each step can be simplified through a discretization of the reflecting surface into pointlike reflectors.


\section{Cassini Thermal Model}


\subsection{Geometric Model}

The first step in this analysis is to build a simplified geometric model of the spacecraft that retains only its main features. This procedure has been validated by a set of test cases performed previously in the analysis of the Pioneer space probes, which gave a good indication that the effect of smaller features does not impact the overall determination of the thermal contribution to the acceleration \cite{Bertolami:2008,Bertolami:2009}. 

\begin{figure}
	\begin{center}
		\epsfxsize=0.75\columnwidth 
		\epsffile{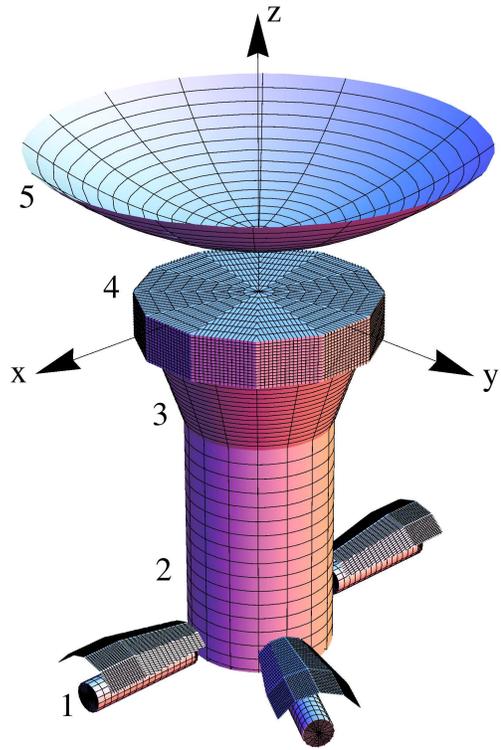}
		\caption{Three-dimensional model of Cassini showing the configuration of the RTGs and their covering structures (1), the spacecraft body composed by a cylindrical lower module (2), a conical upper module (3) and a prismatic main bus (4), and the parabolic high-gain antenna (5).}
		\label{cassini_model}
	\end{center}
\end{figure}

In the case of Cassini, this implies the inclusion of the main antenna dish, the spacecraft body and the three Radioisotope Thermal Generators (RTGs) and respective covers. The main body of the Cassini probe is composed of dodecagonal prism shaped main bus, an upper module with a conic shape and a cylindrical lower module. The three RTGs are attached to the lower model near to its bottom in an asymmetrical configuration. While two of the RTGs are in diametrically opposite positions, the third is at an $120^{\circ}$ angle from one of the latter. Each RTG is covered by an umbrella-like structure composed of eight flat surfaces, arranged as shown in Fig.~\ref{cassini_model}.

Unlike the Pioneer 10 \& 11, the Cassini is not spin stabilized. Instead, it uses an active three-axis stabilization with spin-wheels and thrusters. Due to this fact, the off-axis components of the force are not canceled over time and have to be computed. In any case, judging from the probe's configuration, the component along the $z$-axis should still be dominant; it is also the component for which there are more reliable data for comparison.


\subsection{Order of Magnitude Analysis}
\label{order_mag}

Before embarking on a systematic effort to model the thermal effects on the spacecraft, an analysis of the order of magnitude of the different contributions can provide valuable insight on the task at hand. This analysis helps to identify the most important contributions.

For now, it is enough to consider that the combined power of the RTGs is on the order of $10~{\rm kW}$ and the available electrical power for all the equipment is on the order of $1~{\rm kW}$.

The configuration of the RTGs, each covered with an umbrella-like structure, as depicted in Fig.~\ref{RTG_model}, ensures that a large fraction of the emitted thermal power is absorbed or reflected by the cover, leading to a significant contribution to the thermal force.

From the model of the RTG covers, we find that around $30\%$ of the power emitted by the RTGs, $W_{\rm RTG}$, hits the umbrella-like structures. Due to the shadow cast by this structure, this absorbed radiation does not cancel out radiation propagating in the opposite direction on an otherwise cylindrical radiation field. It is then reasonable to take this value and assume that $30\%$ of the emissions from the RTG are converted into momentum, providing an order of magnitude for the force,
\begin{equation}
	F_{\rm RTG} \sim 0.3 {W_{\rm RTG} \over c} \sim 10^{-5}~{\rm N}.
\end{equation}
Dividing by the spacecraft mass, which for now is assumed to be on the order of $4600~{\rm kg}$ \cite{Bertotti:2003}, we obtain the expected order of magnitude of the thermal acceleration generated by the RTGs
\begin{equation}
	a_{\rm RTG} \sim 2.2 \times 10^{-9}~{\rm m/s^2}.
\end{equation}

When examining the spacecraft body, we can set an upper bound for its contribution, so that it can be compared with the estimates for the effect of the RTGs. To do so, we assume that all the electrical power is dissipated through the bottom wall of the lower compartment. This scenario, albeit simplistic, maximizes the effect of the thermal radiation from the equipment. Under these conditions, we get an upper bound on the force of about
\begin{equation}
	F_{\rm equip} \lesssim {2 \over 3} {W_{\rm elec} \over c} \sim 2.2 \times 10^{-6}~{\rm N},
\end{equation}
and, at most, an acceleration of
\begin{equation}
	a_{\rm equip} \lesssim 4.8 \times 10^{-10}~{\rm m/s^2},
\end{equation}
which is below the estimated effect of the RTGs by a factor of $5$. We stress that this figure clearly overestimates the effects of thermal radiation from electrical power: a more detailed computation will yield a much smaller figure.

This preliminary analysis allows us to conclude that the contribution from the RTGs dominates the thermal acceleration of the Cassini space probe. The obtained order of magnitude also matches the one of the acceleration estimated from the Doppler data.


\subsection{Thermal Radiation Model}

Based on the results of the preceding section, we begin by focusing our attention on the contribution of the RTGs. A significant amount of the radiation emitted from the RTGs is illuminating their covers.

\begin{figure}
	\begin{center}
		\epsfxsize=0.8\columnwidth 
		\epsffile{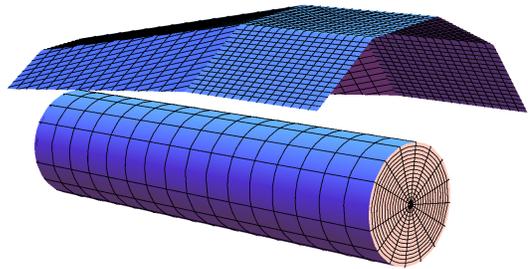}
		\caption{Detail of the geometric model of the umbrella-like structure covering each RTG.}
		\label{RTG_model}
	\end{center}
\end{figure}

The geometric model of the illuminated surface is, in this case, quite realistic, as depicted in Fig.~\ref{RTG_model}. The main issue is, then, to obtain the correct distribution of radiation sources that effectively models the emissions of the RTGs.

In order to discern the sensitivity of the result, we built three different models for the RTG: (i) four isotropic sources uniformly distributed along the centerline, (ii) a cylindrical source along the centerline and (iii) 24 Lambertian sources distributed in four rings of six sources each along the lateral surface of the RTG.

The model with four isotropic sources somewhat underestimates the effect, with a deviation of around $20\%$ relative to the other two. This is due to the fact that it has a significant amount of radiation being emitted laterally in what would still be inside the volume of the RTG. Both other models closely reproduce the cylindrical radiation field, with the results for the power illuminating the RTG cover given by both models within $5\%$ of each other, which is clearly within the accuracy goals set for this study. After analyzing  and comparing the results from these models, we decided to use the cylindrical source configuration, since it is the one that models the radiation field in a more realistic manner.

The first result to be obtained is the fraction of the power emitted that illuminates the covering structure. This figure comes in at $28.3\%$, a part of which is absorbed and the remaining is reflected, depending on the optical properties of the inner surface of the RTG covers. The force computation is made leaving the diffusive and specular reflection coefficients as an open variable to be dealt with later on. 

The contributions to the force are labeled according to Table \ref{contrib_index}, using the spacecraft part numbering from Fig.~\ref{cassini_model}.

\begin{table*}
	\centering
	\caption{Labeling of the considered contributions to the Cassini thermal acceleration.}
	\label{contrib_index}
	\begin{tabular}{l l c}
		\hline
		Emitting surface	& Reflecting surface			& Label \\
		\hline
		\hline
		RTGs				& RTG covers					& $F_{11}$ \\
		Main Bus upper wall	& High-gain antenna back wall	& $F_{45}$ \\
		Lower module bottom wall	& none					& $F_{2}$ \\
		\hline
	\end{tabular}
\end{table*}

We recall that the Cassini has three RTGs positioned in an asymmetrical configuration. For that reason, we first compute the contribution of a single RTG (for simplicity, we start with the one aligned along the $x$-axis). Using the reflection modeling described in Section~\ref{Sec:Phong} and performing the numerical integration, we obtain the force resulting from the emissions of that single RTG
{\setlength\arraycolsep{2pt}
\begin{eqnarray}
	\mathbf{F}_{11} &=& {W_1 \over c} \Big[(-0.0356 - 0.0204 k_{\rm d1} - 0.0466 k_{\rm s1}) \mathbf{e}_x \nonumber \\
	& & + (0.240 + 0.159 k_{\rm d1} + 0.193 k_{\rm s1} )\mathbf{e}_z \Big],
\end{eqnarray}}
where $W_1$ is the power emitted by the RTG and $k_{\rm d1}$ and $k_{\rm s1}$ are the diffusive and specular reflection coefficients of the inner surface of the RTG cover, respectively. In order to obtain the total contribution from the 3 RTGs, we have to add this result to the contribution rotated by $120^{\circ}$ and $180^{\circ}$ around the spacecraft's $z$-axis.

Aside from the RTG contribution, the electric power consumed by the equipment in the spacecraft body also contributes to the thermal acceleration. However, the order of magnitude analysis performed in Section~\ref{order_mag} shows that, at most, it adds up to around $20\%$ of the contribution of the RTGs. Still, the effect of the top and bottom walls can be significant along the $z$-axis and deserves some effort in its determination.

When evaluating the emissions from the top wall of the spacecraft main bus, the main surface illuminated is the back of the parabolic high-gain antenna. The emissions from this surface were modeled through a total of 12 Lambertian sources, each one placed at the centroid of each triangular segment of the dodecagon shaped surface.

Integrating along the antenna, we find that $61.1\% $ of the thermal power emitted from the top wall is hitting the antenna. Assuming that the power is evenly distributed along the surface, the radial components of the source cancels out, leaving only an axial contribution of
\begin{equation}
	\mathbf{F}_{45} = {W_4 \over c} \left(-{2 \over 3} + 0.492 + 0.387 k_{\rm d5} + 0.236 k_{\rm s5} \right)\mathbf{e}_z,
\end{equation}
where $W_4$ is the power emitted from the top wall, $k_{\rm d5}$ is the diffusive reflection coefficient of the antenna and $k_{\rm s5}$ is its specular reflection coefficient.

It should be noted that both the inner wall of the umbrellas and the lower surface of the antenna are modeled with a low shininess constant, $\alpha = 3$ (cf. Eq.~(\ref{specular_reflection})), since these are unpolished surfaces. 

Any amount of power emitted from the bottom wall yields a direct contribution to the acceleration along the $z$-axis, since it does not illuminate any other surface. Considering that it is a Lambertian emitter, if $W_2$ is the power emitted from the bottom wall, then its contribution to the force is
\begin{equation}
	\mathbf{F}_{2} = {2 \over 3} {W_2 \over c} \mathbf{e}_z,
\end{equation}

Given the available information, there is no way to obtain any detailed distribution of the thermal emissions on the lateral walls of the main body of the spacecraft. In any case, such contribution should be very small, since the body has an approximately cylindrical shape and the multilayer insulation blanket tends to even out the temperatures, making the radial radiation field from the space craft body approximately symmetric. For this reason, we focus our attention mainly on the component of the acceleration along the Earth-spacecraft axis, while attempting to get a rough estimate of the other component based entirely on the effect of the RTGs.


\subsection{Power Supply}

The amount of power available on board is of crucial importance for the outcome: the Cassini probe is powered by a set of three large plutonium RTGs; at launch, the RTGs generated around $13~{\rm kW}$ of total thermal power, from which $878~{\rm W}$ of electrical power were produced. Since the plutonium decays with a half-life of $87.7$ years, the total thermal power $W_{\rm Total}$ will decrease at approximately the same rate.

The electrical power generated from the RTGs by a set of thermocouples decreases at a greater rate, due to the decay in the conversion efficiency. This rate of decay can be fitted by an exponential law with a half-life of approximately $21.2$ years \cite{Cooper:2009}. Taking these combined effects into account, the time evolution of the electrical power is given by
\begin{equation}
	\label{ElecPower}
	W_{\rm elec}(t) = 878 e^{-{t \ln 2 \over 87.7}} e^{-{t \ln 2 \over 21.2}}~{\rm W}=  878 e^{-{t \ln 2 \over 17.1}} ~{\rm W},
\end{equation}
with $t$ in years, thus yielding a combined half-life of $17.1$ years.

In order to maintain the overall balance of the spacecraft energy, we assume that the thermal power dissipated at the RTGs results from the difference between total thermal power and the electrical power generated, since the latter will be used to power the array of equipment carried in the spacecraft body,
\begin{equation}
	\label{RTGPower}
	W_{\rm RTG}(t) = W_{\rm total}(t) - W_{\rm elec}(t).
\end{equation}

In this study, we are looking at a very specific period of time, during which the gravitational experiment was performed. As mentioned in the introduction, this corresponds roughly to the month of June 2002, that is, four years and nine months after launch. Given this short time frame, Eq. (\ref{ElecPower}) shows a decrease of only $0.34\%$, so that we can reasonably take the power as constant. Inserting $t=4.75~{\rm years}$ into Eq.~(\ref{ElecPower}) we obtain the reference values for the available power
\begin{eqnarray}
	W_{\rm Total}=12521~{\rm W}~~ &,& ~~W_{\rm elec}=724~{\rm W},\\ \nonumber W_{\rm RTG} &=&11797~{\rm W}.
\end{eqnarray}


\section{Results and Discussion}

\subsection{Baseline Scenarios}

In order to acquire some sensitivity on the influence of the different parameters, prior to a more thorough statistical analysis, we set out a number of scenarios. We consider the spacecraft mass as $m_{\rm Cassini} = 4591~{\rm kg}$ \cite{DiBenedetto:2011}.

The simplest possible scenario, keeping in mind that the RTG contribution is expected to be the dominant one, is to simply consider their effect without reflection on the covers. This means that all power is absorbed and reemitted with the structure at a constant temperature. This results in an acceleration along the $z$-axis,
\begin{equation}
	\mathbf{a}_{\text{Scn 1}} = (-5.09 \mathbf{e}_x -8.81 \mathbf{e}_y + 200 \mathbf{e}_z) \times 10^{-11}~{\rm m/s^2}.
\end{equation}

The next logical step is to include a small amount of reflection from the inner surface of the RTG shades. These structures are covered with a black Kaplan multilayer insulation (MLI), which has a high absorbance of around $90\%$ and also a high emittance of around $0.8$. In terms of the Phong reflection formulation, the translates into a high diffusive reflection coefficient, of around $0.72$ and a specular reflection coefficient of around $0.1$. Tests conducted on the MLI during the development stages of the mission also show that the temperature on the inner layers remains low \cite{Lin:1995}. This also means that there is a small amount of power being transferred to the RTG cover's inner structure, precluding any significant power transfer to the main body through heat conduction from the RTG shades.

Translating this to our model, we first consider as a conservative estimate, a diffusive reflection coefficient of $0.4$ and a specular reflection coefficient of $0.1$. These conditions yield an acceleration of
\begin{equation}
	\mathbf{a}_{\text{Scn 2}} = (-6.87 \mathbf{e}_x -11.9 \mathbf{e}_y + 269 \mathbf{e}_z) \times 10^{-11}~{\rm m/s^2}.
	\label{Scenario2}
\end{equation}

To obtain an upper bound for the RTG contribution, we set the reflectivity coefficients at double the previous scenario, which would mean a total reflection of the thermal power irradiating the inner surface of the RTG covers. This hypothesis yields an acceleration of
\begin{equation}
	\mathbf{a}_{\text{Scn 3}} = (-8.65 \mathbf{e}_x -15.0  \mathbf{e}_y + 337 \mathbf{e}_z) \times 10^{-11}~{\rm m/s^2}.
\end{equation}

If we add to the previous conditions, the upper bound for the contribution from the electrical equipment, meaning that all the power would be dissipated through the lower wall, we get a slightly larger acceleration on the $z$-axis,
\begin{equation}
	\mathbf{a}_{\text{Scn 4}} = (-8.65 \mathbf{e}_x -15.0 \mathbf{e}_y + 372 \mathbf{e}_z) \times 10^{-11}~{\rm m/s^2}.
\end{equation}
This scenario gives us the upper limit for the overall acceleration given by our model.

A more reasonable scenario, is to take the second one considered above, using the reflection coefficients of $0.4$ and $0.1$, and add to it a contribution from the spacecraft body that assumes that power is dissipated uniformly through all the surfaces. The MLI blanket covering the spacecraft body has the effect of evening out major temperature differences along the probe's structure, making this hypothesis reasonable. This scenario yields a small increase in the $z$ component of the acceleration relative to Eq.~(\ref{Scenario2}),
\begin{equation}
	\mathbf{a}_{\text{Scn 5}} = (-6.87 \mathbf{e}_x -11.9 \mathbf{e}_y + 272 \mathbf{e}_z) \times 10^{-11}~{\rm m/s^2}.
\end{equation}
This last set of hypotheses represents the baseline for the parametric study that follows in the next section. Notwithstanding, we can already point out that the $z$ component is remarkably close to the value of $3 \times 10^{-9}~{\rm m/s^2}$, reported through the Doppler analysis \cite{Bertotti:2003}.

The results from all the considered scenarios are summarized in Table~\ref{scenarios}.

\begin{table*}
	\centering
	\caption{Summary of assumptions and results from baseline scenarios, where $W_{\text{bottom}}$ and $W_{\text{top}}$ are the emitted powers from the bottom wall of the lower module and the top wall of the main bus, respectively; $k_{\rm d,umbr}$ and $k_{\rm s,umbr}$ are the RTG umbrella reflection coefficients, and $a_x$, $a_y$ and $a_z$ are the components of the resulting thermal acceleration.}
	\label{scenarios}
	\begin{tabular}{l c c c c c c c}
		\hline
		Scenario & $W_{\text{bottom}}$ & $W_{\text{top}}$	& $k_{\rm d,umbr}$ &  $k_{\rm s,umbr}$ & $a_x$  & $a_y$  & $a_z$ \\
		~ & $(W)$ & $(W)$ & ~ & ~ & $(10^{-11}{\rm m/s^2})$  & $(10^{-11}{\rm m/s^2})$ & $(10^{-11}{\rm m/s^2})$ \\
		\hline
		\hline
		1. RTGs, no reflection			& $0$	& $0$	& $0$	& $0$	& $-5.09$ 	& $-8.81$	& $200$ \\
		2. RTGs, low reflection			& $0$	& $0$	& $0.4$	& $0.1$	& $-6.87$ 	& $-11.9$	& $269$ \\
		3. RTGs, high reflection		& $0$	& $0$	& $0.8$	& $0.2$	& $-8.65$ 	& $-15.0$	& $337$ \\
		4. Upper bound					& $724$	& $0$	& $0.8$	& $0.2$	& $-8.65$ 	& $-15.0$	& $372$ \\
		5. Scenario 2 + body uniform temp.	& $47.2$& $121$	& $0.4$	& $0.1$	& $-6.87$ 	& $-11.9$	& $272$ \\
		\hline
	\end{tabular}
\end{table*}

The off-axis components, however, remain about one order of magnitude below the values reported in Ref.~\onlinecite{Bertotti:2003} --- although the latter are quite unreliable, as the authors themselves point out. Furthermore, those values are presented relative to the orbital plane, whereas the results of the thermal analysis correspond to the spacecraft reference frame. 

One could speculate that this difference is due to the rotation between a reference with the $z$ axis along the axis of the high-gain antenna and one with the $z$ axis on the orbital plane. A simple calculation, hypothesizing that the antenna is pointing directly towards the Earth can be performed using data from the \emph{Cassini, Galileo, and Voyager ephemeris tool} \cite{CassiniEphemeris}: during the solar conjunction experiment, the angle between the two reference frames would be between $1.6^{\circ}$ and $1.8^{\circ}$. The projection of the $z$ component of the acceleration on the spacecraft frame on a direction orthogonal to the orbital plane would result in an acceleration component close to $10^{-10}~{\rm m/s^2}$, which would agree with the order of magnitude of the Doppler measurements.

In the absence of more complete information on the methods used to obtain the Doppler estimates and the spacecraft orientation during the time of the experiment, it is not possible to make any definite assertions about the off-axis components of the acceleration.

\subsection{Parametric Analysis}

We now proceed to the statistical analysis based on a Monte Carlo simulation. We focus this analysis only on the $z$ component of the acceleration, since there is not enough information to properly constrain the relevant parameters for the off-axis component and, as discussed above, the results would not be reliable enough to draw any conclusions.

In the Monte Carlo method, a large number of random values associated with a statistical distribution are generated for each of the relevant parameters that influence the final result. This type of analysis turned out to be quite valuable for the Pioneer anomaly problem \cite{Francisco:2012}.

The reflection coefficients of the RTG umbrellas can be reasonably constrained, as discussed in the preceding section, since we have some reliable data on the material and its properties \cite{Lin:1995}. This enables us to use a Normal distribution for the diffusive reflection coefficient, centered at a conservative estimate of $0.6$ and with a $\sigma$ of $0.1$, allowing for a variation between $0.4$ and $0.8$ within $2\sigma$. For all other parameters, we use uniform distributions with a reasonably wide interval in order to account for the lack of accurate and reliable data. The assumptions used to generate the random values for the simulation are outline in Table~\ref{MCparameters}.

\begin{table*}
	\centering
	\caption{Assumptions used to generate input parameter values for Monte Carlo simulation.}
	\label{MCparameters}
	\begin{tabular}{l l l}
		\hline
		Parameter		& Symbol 	& Distribution \\
		\hline
		\hline
		Diffusive reflection coeff. of umbrella		& $k_{\rm d,umbr}$ & Normal distribution with $\mu = 0.6$ and $\sigma= 0.1$ \\
		Specular reflection coeff. of umbrella		& $k_{\rm s,umbr}$ & Uniform distribution with interval $[0,0.2]$ \\
		Diffusive reflection coeff. of antenna back	& $k_{\rm d,ant}$ &  Uniform distribution with interval $[0,0.6]$ \\
		Specular reflection coeff. of antenna back	& $k_{\rm s,ant}$ &  Uniform distribution with interval $[0,0.2]$ \\
		Power emitted from top of main bus			& $W_{\text{top}}$ & Uniform distribution with interval $[0,94.4]~{\rm W}$ \\
		Power emitted from bottom of lower module	& $W_{\text{bottom}}$ & Uniform distribution with interval $[0,242]~{\rm W}$ \\
		\hline
	\end{tabular}
\end{table*}

\begin{figure}
	\begin{center}
		\epsfxsize=\columnwidth 
		\epsffile{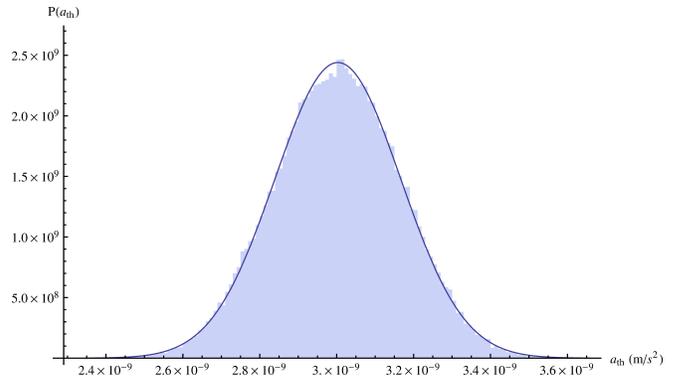}
		\caption{Probability density distribution resulting from the Monte Carlo simulation of the thermal acceleration along the $z$-axis, with the normal distribution with the same mean and standard deviation superimposed.}
		\label{Histogram}
	\end{center}
\end{figure}

Running a simulation with $10^5$ iterations, we obtain the probability density distribution depicted in Fig~\ref{Histogram}. The distribution is approximately normal. The mean of the resulting distribution is $3.01 \times 10^{-9}~{\rm m/s^2}$, with a standard deviation of $1.63 \times 10^{-10}~{\rm m/s^2}$. The acceleration along the Earth-spacecraft axis, with an uncertainty interval of $2\sigma$, is
\begin{equation}
	(a_{\text{Cassini}})_z = (3.01 \pm 0.33) \times 10^{-9}~{\rm m/s^2}.
\end{equation}

From this analysis, we can conclude that the value for the thermal acceleration given by this model of the Cassini spacecraft is in agreement with the value obtained from the Doppler data, up to a $95\%$ probability level.


\section{Conclusions}

The results found in this study for the thermally generated acceleration of the Cassini space probe during its solar conjunction experiment significantly reinforce our confidence in the method first developed to account for the Pioneer anomaly \cite{Bertolami:2008,Bertolami:2009,Francisco:2012}. The adaptability of this approach allowed its application to an entirely new problem, with a different geometry, material properties and set of hypotheses, upholding the transparency and the simplicity of the method.

Clearly, some open questions still remain. The off-axis components of the acceleration are still poorly known. More detailed  information about the internal power consumption and the attitude of the probe would be needed to properly address this issue. However, the result for the main component of the acceleration, along the probe's $z$ axis, gives a very compelling result that closely agrees with the estimates of the non-gravitational acceleration presented in Ref.~\onlinecite{Bertotti:2003}.

The results presented in this paper significantly boost the confidence in one of the most accurate experiments ever performed to test General Relativity.\\[1cm]


\section*{Acknowledgments}

The work of FF is sponsored by the FCT -- Funda\c{c}\~{a}o para a Ci\^{e}ncia e Tecnologia (Portuguese Agency), under the grant BD 66189/2009.

The work of OB and JP is partially supported by the FCT grant PTDC/FIS/111362/2009.


\bibliography{cassini}

\end{document}